\documentclass[aps,prc,letterpaper,11pt,twoside,tightenlines,nofootinbib,showpacs,preprint]{revtex4}
\usepackage{graphicx}
\usepackage[sort&compress]{natbib}
\begin{document}

%%%   New Definitions
%\newcommand{\eg}{{\it e.g.}}
%\newcommand{\etal}{{\it et. al.}}
%\newcommand{\ie}{{\it i.e.}}
\newcommand{\be}{\begin{equation}}
\newcommand{\ee}{\end{equation}}
\newcommand{\bea}{\begin{eqnarray}}
\newcommand{\eea}{\end{eqnarray}}
\newcommand{\bef}{\begin{figure}}
\newcommand{\eef}{\end{figure}}
\newcommand{\bce}{\begin{center}}
\newcommand{\ece}{\end{center}}

\newcommand{\bfn}{{\bm {n}}}
\newcommand{\bfJ}{{\bm {J}}}

\newcommand{\pion}{{\pi}}
\newcommand{\kaon}{{K}}
\newcommand{\kaonb}{{\bar{K}}}
\newcommand{\kaonS}{{K^*}}
\newcommand{\kaonSb}{{\bar{K}^*}}
\newcommand{\D}{{D}}
\newcommand{\Db}{{\bar{D}}}
\newcommand{\Ds}{{D_s}}
\newcommand{\Dsb}{{\bar{D}_s}}
\newcommand{\DS}{{D^*}}
\newcommand{\DSb}{{\bar{D}^*}}
\newcommand{\DsS}{{D_s^*}}
\newcommand{\DsSb}{{\bar{D}_s^*}}
\newcommand{\etap}{{\eta^\prime}}
\newcommand{\etac}{{\eta_c}}
\newcommand{\Jpsi}{{\psi}}
\newcommand{\Nucleon}{{N}}
\newcommand{\Lambdac}{{\Lambda_c}}
\newcommand{\SigmaS}{{\Sigma^*}}
\newcommand{\Sigmac}{{\Sigma_c}}
\newcommand{\SigmacS}{{\Sigma_c^*}}
\newcommand{\Cascada}{{\Xi}}
\newcommand{\Cascadac}{{\Xi_c}}
\newcommand{\CascadaS}{{\Xi^*}}
\newcommand{\CascadacS}{{\Xi_c^*}}
\newcommand{\Cascadacc}{{\Xi_{cc}}}
\newcommand{\CascadaccS}{{\Xi_{cc}^*}}
\newcommand{\Cascadacp}{{\Xi_c^\prime}}

\newcommand{\Omegac}{{\Omega_c}}
\newcommand{\Omegacc}{{\Omega_{cc}}}
\newcommand{\OmegacS}{{\Omega_c^*}}
\newcommand{\OmegaccS}{{\Omega_{cc}^*}}
\newcommand{\Omegaccc}{{\Omega_{ccc}}}

\newcommand{\Qur}{{Q^\dagger_{u\uparrow}}}
\newcommand{\Quj}{{Q^\dagger_{u\downarrow}}}
\newcommand{\Qdr}{{Q^\dagger_{d\uparrow}}}
\newcommand{\Qdj}{{Q^\dagger_{d\downarrow}}}
\newcommand{\Qsr}{{Q^\dagger_{s\uparrow}}}
\newcommand{\Qsj}{{Q^\dagger_{s\downarrow}}}
\newcommand{\Qcr}{{Q^\dagger_{c\uparrow}}}
\newcommand{\Qcj}{{Q^\dagger_{c\downarrow}}}

\newcommand{\Qubr}{{Q^\dagger_{\bar{u}\uparrow}}}
\newcommand{\Qubj}{{Q^\dagger_{\bar{u}\downarrow}}}
\newcommand{\Qdbr}{{Q^\dagger_{\bar{d}\uparrow}}}
\newcommand{\Qdbj}{{Q^\dagger_{\bar{d}\downarrow}}}
\newcommand{\Qsbr}{{Q^\dagger_{\bar{s}\uparrow}}}
\newcommand{\Qsbj}{{Q^\dagger_{\bar{s}\downarrow}}}
\newcommand{\Qcbr}{{Q^\dagger_{\bar{c}\uparrow}}}
\newcommand{\Qcbj}{{Q^\dagger_{\bar{c}\downarrow}}}

\newcommand{\uno}{{\bf 1}}
\newcommand{\dos}{{\bf 2}}
\newcommand{\tres}{{\bf 3}}
\newcommand{\cuatro}{{\bf 4}}
\newcommand{\ocho}{{\bf 8}}
\newcommand{\diez}{{\bf 10}}
\newcommand{\quince}{{\bf 15}}
\newcommand{\dieciseis}{{\bf 16}}
\newcommand{\veinte}{{\bf 20}}
\newcommand{\veintisiete}{{\bf 27}}
\newcommand{\treintaycinco}{{\bf 35}}
\newcommand{\cincuentayseis}{{\bf 56}}
\newcommand{\sesentaytres}{{\bf 63}}
\newcommand{\cientoveinte}{{\bf 120}}
\newcommand{\cientosesentayocho}{{\bf 168}}
\newcommand{\setecientosveinte}{{\bf 720}}
\newcommand{\novecientoscuarentaycinco}{{\bf 945}}
\newcommand{\mildoscientostreintaydos}{{\bf 1232}}
\newcommand{\dosmilquinientosveinte}{{\bf 2520}}
\newcommand{\cuatromilsetecientoscicuentaydos}{{\bf 4752}}
\newcommand{\trecemilcientocuatro}{{\bf 13104}}

\title{The properties of  $D$ and $D^*$ mesons in the nuclear medium}

\author{ L. Tolos$^1$, C. Garcia-Recio$^2$ and J. Nieves$^3$\\
$^1$ Theory Group. KVI. University of Groningen, \\
Zernikelaan 25, 9747 AA Groningen, The Netherlands \\
$^2$ Departamento de F{\'\i}sica At\'omica, Molecular y Nuclear,\\ 
Universidad de Granada, E-18071 Granada, Spain\\
$^3$ Instituto de F{\'\i}sica Corpuscular (centro mixto CSIC-UV)\\
Institutos de Investigaci\'on de Paterna, Aptdo. 22085, 46071, Valencia, Spain
}

\begin{abstract}

We study the properties of $D$ and $D^*$ mesons in nuclear matter
within a simultaneous self-consistent coupled-channel unitary approach
that implements heavy-quark symmetry. The in-medium solution accounts
for Pauli blocking effects, and for the $D$ and $D^*$ self-energies in a
self-consistent manner. We pay a special attention to the
renormalization of the intermediate propagators in the medium beyond
the usual cutoff scheme. We analyze the behavior in the nuclear
medium of the rich spectrum of dynamically-generated baryonic
resonances in the $C=1$ and $S=0$ sector, and their influence in the
self-energy and, hence, the spectral function of the $D$ and $D^*$
mesons. The $D$ meson quasiparticle peak mixes with
$\Sigma_c(2823)N^{-1}$ and $\Sigma_c(2868)N^{-1}$ states while the
$\Lambda_c(2595)N^{-1}$ mode is present in the low-energy tail of the
spectral function. The $D^*$ spectral function incorporates $J=3/2$
resonances, and $\Sigma_c(2902)N^{-1}$ and $\Lambda_c(2941)N^{-1}$
fully determine the behavior of $D^*$ meson spectral function at the
quasiparticle peak. As density increases, these resonant-hole modes
tend to smear out and the spectral functions get broad. We also obtain
the $D$ and $D^*$ scattering lengths, and optical potentials for
different density regimes. The $D$ meson potential stays
attractive while the $D^*$ meson one is repulsive with increasing densities up to twice that of the normal nuclear matter. Compared to
previous in-medium SU(4) models, we obtain similar values for the real
part of the $D$ meson potential but much smaller imaginary parts. This
result can have important implications for the observation of
$D^0$-nucleus bound states.

\end{abstract}

\pacs{11.10.St, 12.38.Lg, 14.20.Lq, 14.40.Lb, 21.65.-f}

\vspace{1cm}

\date{\today}

\maketitle

\section{Introduction}
\label{sec:intro}

The interest on the properties of open and hidden charmed mesons was
triggered more than 20 years ago in the context of relativistic
nucleus-nucleus collisions in connection to the charmonium suppression
\cite{matsui} as a probe for the formation of Quark-Gluon Plasma (QGP). The
experimental programme in hadronic physics of the future FAIR facility
at GSI \cite{fair} will move from the light quark sector to the heavy
one and will face new challenges where charm plays a dominant role. In
particular, a large part of the PANDA physics programme will be
devoted to charmonium spectroscopy. Moreover, the CBM experiment will
extend the GSI programme for in-medium modification of hadrons in the
light quark sector, and provide first insight into the charm-nucleus
interaction.

The primary theoretical effort is to understand the interaction
between hadrons with the charm degree of freedom. Charmed baryonic
resonances have received recently a lot of attention motivated by the
discovery of quite a few new states by the CLEO, Belle and BABAR
collaborations
\cite{2880-Artuso:2000xy,Mizuk:2004yu,Chistov:2006zj,Aubert:2006je,2940-Aubert:2006sp,2880-Abe:2006rz}. Whether
those resonances have the usual $qqq$ structure or qualify better as
being dynamically generated via meson-baryon scattering processes is a
matter of strong interest. In fact, the unitarization, in
coupled-channels, of the chiral perturbation amplitudes for scattering
of $0^-$ octet Goldstone bosons off baryons of the nucleon $1/2^+$
octet has proven to be quite successful in the charmless sector
\cite{Kai95,Kaiser:1995cy,OR98,kr98,Nacher:1999vg,Meissner:1999vr,Oller:2000fj,
  Jido:2003cb,Nieves:2001wt,Inoue:2001ip,kolo,Garcia-Recio:2002td,ORB02,Ramos:2002xh,Tolos:2000fj,GarciaRecio:2003ks,Oller:2005ig,Borasoy:2005ie,Borasoy:2006sr,Hyodo:2008xr}.
The modification of the various meson-baryon amplitudes for the case
of finite temperature and/or nuclear density has also attracted a lot
of attention and has been carefully discussed~\cite{Ramos:1999ku,
Tolos:2002ud, Tolos:2005jg,Tolos:2006ny, Tolos:2008di,
Cabrera:2009qr}.

The extension to the charm sector of the unitarized meson-baryon method
was attempted in a first exploratory work in
Ref.~\cite{tolos-schaffne-mishra}, where the free space amplitudes
were constructed from a set of separable coupled-channel interactions
obtained from chirally motivated lagrangians upon replacing the $s$
quark by the $c$ quark. A different approach resulting from the
scattering of Goldstone bosons off the ground state $1/2^+$ charmed
baryons was  pursued in \cite{kolo}, but the substantial improvement
in constructing the meson-baryon interaction in the charm sector came
from exploiting the universal vector-meson coupling hypothesis to
break the SU(4) symmetry \cite{hofmann}. The $t$-channel exchange of
vector mesons (TVME) between pseudoscalar mesons and baryons preserved
chiral symmetry in the light meson sector keeping the
Weinberg-Tomozawa (WT) type of interaction. An extension to $d$-wave
$J=3/2^-$ resonances was developed in \cite{hofmann2}, while some
modifications over the model of Ref.~\cite{hofmann} were implemented
in Ref.~\cite{angels-mizutani}, both in the kernel and in the
renormalization scheme. More recently, there have been attempts to
construct the $DN$ and $\bar DN$ interaction by incorporating the
charm degree of freedom in the SU(3) meson-exchange model of the
J\"ulich group \cite{haide1,haide2}.

Nuclear medium modifications were then incorporated in order to study
the properties of charmed mesons in nuclear matter, and the influence
of those modifications in the charmonium production rhythm at finite
baryon densities. Possible variation of this rhythm might indicate the
formation of the QGP phase of QCD at high densities.  Previous
works based on mean-field approaches provided important mass shifts
for the $D$ and $\bar D$ meson masses
\cite{tsushima,sibirtsev,hayashigaki,mishra}.  Some of those models
have been recently revised \cite{hilger,mishra08}. However, the
spectral features of $D$ mesons in symmetric nuclear matter were
obtained for the first time in the exploratory work of
Ref.~\cite{tolos-schaffne-mishra}, while finite temperature effects
were incorporated later on in Ref.~\cite{tolos-schaffne-stoecker}.
Afterwards, within the SU(4) TVME model of Ref.~\cite{hofmann}, the
properties of the ($D$,$\bar D$) and ($D_s$, $\bar D_s$) mesons were
analyzed in ~\cite{lutz}, and in
~\cite{angels-mizutani,tolos-angels-mizutani}. In this latter
reference, the kernel and the renormalization scheme employed in
Ref.~\cite{hofmann} were modified.

However, those SU(4) TVME inspired models are not consistent with
heavy-quark symmetry (HQS), which is a proper QCD spin-flavor symmetry
that appears when the quark masses, such as the charm mass, become
larger than the typical confinement scale. As a consequence of this
symmetry, the spin interactions vanish for infinitely massive
quarks. Thus, heavy hadrons come in doublets (if the spin of the light
degrees of freedom is not zero), which are degenerated in the infinite
quark-mass limit. And this is the case for the $D$ meson and its
vector partner, the $D^*$ meson.

In fact, the incorporation of vector mesons into the coupled-channel
picture has been pursued very recently in the strange sector. On one
hand, vector mesons have been incorporated within the hidden-gauge
formalism. Within this scheme, a broad spectrum of new resonant
meson-baryon states have been generated~\cite{Sarkar:2009kx,
Oset:2009sw,Oset:2009vf}. On the other hand, the WT meson-baryon
chiral Lagrangian has been also extended to account for vector meson
degrees of freedom by means of a scheme that starts from a SU(6)
spin-flavor symmetry Lagrangian and that incorporates some symmetry
breaking corrections determined by physical masses and meson decay
constants~\cite{Garcia-Recio:2005hy,GarciaRecio:2006wb,QNP06,Toki:2007ab}. The
corresponding Bethe-Salpeter equation reproduces the
previous SU(3)-flavor WT results for the lowest-lying $s$- and
$d$-wave, negative parity baryon resonances and gives new information
on more massive states, as for example the $\Lambda(1800)$ or
$\Lambda(2325)$ resonances. The extension of this scheme to four flavors,
incorporating the charm degree of freedom, was carried out in
Ref.~\cite{GarciaRecio:2008dp} and it automatically incorporates HQS in the charm
sector improving in this respect on the  SU(4) TVME models, since $D$ and
$D^*$ mesons are thus consistently treated. One of the distinctive
differences of this approach with respect to those that built in the SU(4)
TVME model can be drawn in the wave function content of the 
resonances.  Thus, for instance, the dynamics of the lowest lying resonance
$\Lambda_c(2595)$ is completely dominated by the $DN$ channel in the 
SU(4) TVME model of Ref.~\cite{hofmann}, while it turns out be largely
a $D^*N$ state within the SU(8) scheme of Ref.~\cite{GarciaRecio:2008dp}. Such
differences might have also a direct influence on the dynamics at
finite densities. 

Thus in this work, we aim to investigate the nuclear medium effects in
hadronic systems with charm one ($C=1$) and no strangeness ($S=0$) within
the SU(8) model derived in \cite{GarciaRecio:2008dp}. In particular, we study the
dynamically-generated baryonic resonances in the free space as well as
in the nuclear medium in order to analyze how the masses and widths
are modified with density. We also study the $D$ and $D^*$ mesons
self-energies in the nuclear medium, calculating their spectral
functions for a variety of densities and
the corresponding optical potentials. A novelty of our work is that we
simultaneously obtain, in a self-consistent manner, the $D$ and $D^*$ meson
self-energies. We then compare our results with the previous ones
obtained within SU(4) TVME schemes and other more simple models
\cite{tolos-schaffne-mishra,tolos-schaffne-stoecker,lutz,angels-mizutani,tolos-angels-mizutani},
paying also an special attention to the regularization of the intermediate
propagators in the medium beyond the cutoff method. The nuclear medium
effects for hadronic scattering amplitudes and for hadrons propagators
are of interest for the understanding and correct interpretation of the data
obtained in heavy-ion collisions where the high nuclear densities
reached can substantially change the properties of the involved hadrons.

To end this introduction, and before we start deriving the $D$ and
 $D^*$ self-energies inside of a nuclear medium, we would like to make
 a general reflexion to better situate this work. Though the model of
 Ref.~\cite{GarciaRecio:2008dp} is possibly the best one existing in the literature
 for describing the free space $C=1, S=0$ meson--baryon elastic
 scattering at low energies\footnote{As outlined above, it provides a
 scheme for four flavors and for pseudoscalar and vector mesons which
 reduces to the WT interaction in the sector where Goldstone bosons
 are involved, and that incorporates HQS in the charm sector.}, it
 does not provide, as it is the case for the rest of the available
 models, the correct analytical properties of the scattering
 matrix, including $s,t$ and $u-$cuts and proper crossing
 symmetry. Only the real theory, presumably QCD, could do
 that. However, one should bear in mind that deficiencies of the 
 free space model would definitely affect to results that will be
 presented here for amplitudes embedded in cold nuclear matter. Since
 we aim at describing resonances, it is important to use a model
 consistent with the unitarity cut as that of Ref.~\cite{GarciaRecio:2008dp}, while
 for rest of analytical properties one hopes to be partially taken
 into account thanks to low energy constants (subtraction
 constants...). To ignore a cut that would cover totally/partially the
 studied resonance region is not so much important\footnote{For
 instance, the $t-$channel $\rho$ exchange cut for the elastic $\bar
 D^* \Sigma^*_c$ amplitude goes from about 0.53~GeV to 4.45~GeV. 
 Obviously, one does not need to consider explicitly this cut when
 studying the elastic $\pi N$ scattering in the region of the
 $\Delta(1232)$ resonance.}, and what  is more relevant is the
 proximity of a resonance to a branching point, where the amplitude
 might vary more rapidly, together with the strength of the coupling
 of the resonance to the given channel. The free space model of
 Ref.~\cite{GarciaRecio:2008dp} only properly accounts for the two body unitarity
 cut, and it thus suffers  from some limitations. As a matter of
 example, this model does not account for the dynamics of an
 intermediate nucleon in the $u-$channel diagram $DN \to \Lambda_c
 \rho$, involving two $p-$wave couplings ($DN\Lambda_c$ and $NN\rho$),
 that gives rise to a cut extending from 1.55 to 2.74~GeV.  The lowest
 energy branch point should not suppose a serious problem. On the
 other hand, $\Lambda_c \rho$ selects isospin $I=1$, and if one looks
 at Tables VI and IX of Ref. [57], one sees no resonances excessively
 close to the branch point 2.74~GeV, with a dominant coupling to that
 channel.  Thus, though some effects will exist, we do not expect
 drastic changes that could lead us to think that the interaction
 model of Ref.~\cite{GarciaRecio:2008dp} is unrealistic, specially close to
 threshold where the $p-$wave couplings will be negligible.

 Other contributions no considered in the model of Ref.~\cite{GarciaRecio:2008dp}
 are those driven by the exchange of a pion in the $t-$channel. Among
 all of them that contributing to the $DN\to D^*N$ amplitude could be
 quite relevant, since it might induce a pronounced energy dependence,
 difficult to account by means of the low energy constants of the
 model\footnote{Previous models~\cite{tolos-schaffne-mishra, kolo,
 hofmann} also ignored this contribution because in these
 works $D^*$-degrees of freedom were not explicitly taken into account
 in the external legs. But, as we have commented, ignoring such
 degrees of freedom was totally unjustified.}.  This contribution
 involves again two $p-$wave vertices, which would vanish at the
 $D^*N$ threshold. Below threshold, its contribution would heavily
 depend on the adopted form factors to account for the off-shellness.
 This $t-$pion-exchange contribution can be safely neglected at
 threshold. Above threshold, and because the $D$ and $D^*$ meson
 masses are quite similar, the exchanged virtual pion will carry a
 small energy and therefore in good approximation $q^2 \sim
 -\vec{q}^{\,2}$. Having in mind the existence of two $p-$wave
 vertices in the diagram, each of them  proportional to the virtual pion
 momentum, we expect a large cancellation of these with the pion
 propagator (neglecting the pion mass), which will significantly
 reduce the energy dependence of this contribution.

 From the above discussion, it is clear that the predictions of the
 model of Ref.~\cite{GarciaRecio:2008dp} far from threshold are likely subject to
 large uncertainties due to the contributions mentioned above (pion
 exchanges in the $t-$channel, $u-$cut contributions), and possibly
 other mechanisms.  This is the reason, why in this work
 we have only computed properties of the in medium amplitudes
 close to threshold.

\section{Formalism: the $DN$ and $D^*N$ interaction in nuclear matter}

We will calculate the self-energy of the $D$ and $D^*$ mesons in
nuclear matter from a self-consistent calculation in coupled channels
that treats the heavy pseudoscalar and vector mesons on equal footing,
as required by HQS. To incorporate HQS to the meson-baryon interaction
we extend the WT meson-baryon lagrangian to the SU(8) spin-flavor
symmetry group \cite{GarciaRecio:2008dp}. We start from the traditional three flavor
WT Lagrangian, which is not just SU(3) symmetric but also chiral
$\left(SU_L(3)\otimes SU_R(3)\right)$ invariant. Symbolically, up to
an overall constant, the WT interaction is \be {\cal L_{\rm WT}}= {\rm
Tr} ( [M^\dagger, M][B^\dagger, B]) \,, \ee where mesons ($M$) and
baryons ($B$) fall in the SU(3) representation $\ocho$, which is the
adjoint representation. The commutator indicates a $t-$channel
coupling to the $\ocho_a$ (antisymmetric) representation. For the
SU(8) spin--flavor symmetry, the mesons $M$ fall now in the
$\sesentaytres$ (adjoint representation) and the baryons $B$ are found
in the $\cientoveinte$, which is fully symmetric. The group reductions
\begin{eqnarray}
\sesentaytres\otimes\sesentaytres &=& \uno\oplus\sesentaytres_s
\oplus\sesentaytres_a\oplus\setecientosveinte\oplus\novecientoscuarentaycinco\oplus\novecientoscuarentaycinco^*\oplus\mildoscientostreintaydos
\nonumber \\
\cientoveinte\otimes\cientoveinte^* &=&
\uno\oplus\sesentaytres\oplus\mildoscientostreintaydos\oplus\trecemilcientocuatro 
\end{eqnarray}
lead to a total of four different $t-$channel SU(8) singlet couplings,
that can be used to construct $s$-wave meson-baryon interactions
\begin{eqnarray}
\left((M^\dagger\otimes M)_{\uno}\otimes
(B^\dagger\otimes B)_{\uno}\right)_{\uno}, &&
\left((M^\dagger\otimes M)_{\sesentaytres_a}\otimes
(B^\dagger\otimes B)_{\sesentaytres}\right)_{\uno},
\nonumber \\
\left((M^\dagger\otimes M)_{\sesentaytres_s}\otimes
(B^\dagger\otimes B)_{\sesentaytres}\right)_{\uno}, &&
\left((M^\dagger\otimes M)_{\mildoscientostreintaydos}\otimes
(B^\dagger\otimes B)_{\mildoscientostreintaydos}\right)_{\uno}.
\label{eq:coupl}
\end{eqnarray}

To ensure that the SU(8) amplitudes will reduce to those deduced from
the SU(3) WT Lagrangian in the
$(\ocho_\uno)$meson--$(\ocho_\dos)$baryon subspace (denoting the SU(3)
multiplets of dimensionality $\bf n$ and spin $J$ by ${\bf
n}_{\dos{\bf J}+\uno}$), we set all the couplings in
Eq.~(\ref{eq:coupl}) to be zero except for
\begin{equation}
{\cal L_{\rm WT}^{\rm SU(8)}}=
\left((M^\dagger\otimes M)_{\sesentaytres_a}\otimes
(B^\dagger\otimes B)_{\sesentaytres}\right)_{\uno}  \ ,
\label{eq:NOcoupl}
\end{equation}
which is the natural and unique SU(8) extension of the usual SU(3) WT
Lagrangian. To compute the matrix elements of the SU(8)
WT interaction,  ${\cal L_{\rm WT}^{\rm SU(8)}}$, we use quark model constructions of hadrons 
with field theoretical methods to express everything
in tensor representations as
described in  Appendix A of Ref.~\cite{GarciaRecio:2008dp}.
Thus, we get the tree level
amplitudes (we use the convention $V=-{\cal L}$):
\begin{equation}
V^{IJSC}_{ab}(\sqrt{s})=
D^{IJSC}_{ab} \frac{\sqrt{s}-M}{2\,f^2}
\left(\sqrt{\frac{E+M}{2M}}\right)^2 \ ,
\label{eq:vsu8}
\end{equation}
where the last factor is due to the spinor normalization convention:
$\bar u u =\bar v v =1$, as in Refs.~\cite{ORB02,Jido:2003cb}.  In the
above expression $IJSC$ are the meson--baryon isospin, total angular
momentum, strangeness and charm quantum numbers, $M~(E)$ the common
mass (CM energy) of the baryons placed in the $\cientoveinte$ SU(8)
representation, and $D^{IJSC}$ a matrix in the coupled channel
space (see Ref.~\cite{GarciaRecio:2008dp}).

However, the SU(8) spin-flavor is strongly broken in nature. The
breaking of SU(8) is twofold. On one hand, we take into account mass
breaking effects by adopting the physical hadron masses in the tree
level interactions of Eq.~(\ref{eq:vsu8}) and in the evaluation of the
kinematical thresholds of different channels. On the other hand, we
consider the difference between the weak non-charmed and charmed
pseudoscalar and vector meson decay constants. Then, our tree level
amplitudes now read
\begin{equation}
V^{IJSC}_{ab}(\sqrt{s})= D^{IJSC}_{ab}
\frac{2\sqrt{s}-M_a-M_b}{4\,f_a f_b} \sqrt{\frac{E_a+M_a}{2M_a}}
\sqrt{\frac{E_b+M_b}{2M_b}} \,, \label{eq:vsu8break}
\end{equation}
where $M_a$ ($M_b$) and $E_a$ ($E_b$) are, respectively, the mass and
the CM energy of the baryon in the $a$ ($b$) channel.We focus in the
non-strange ($S=0$) and singly charmed ($C=1$) sector, where the $DN$ and
$D^*N$ are embedded. In particular, we look at $I=0$ and $I=1$
channels for $J=1/2$ and $J=3/2$. The channels involved in the
coupled-channel calculation are given in Table~\ref{table1},
\begin{table}[htb]
\caption{For each isospin $IJ$ sector, all the involved baryon-meson
channels are compiled, with their mass thresholds, $M+m$, in MeV
shown below.}
\begin{center}
{$I=0$, $J=1/2$}
\end{center}
\begin{tabular}{@{\extracolsep{1mm}}cccccccccccccccccccc}
\hline
 $ \Sigmac \pion $ &  $ \Nucleon \D 
  $ &  $ \Lambdac \eta $ 
 &  $ \Nucleon \DS $ &  $ \Cascadac 
  \kaon $ &  $ \Lambdac \omega $ 
 &  $ \Cascadacp \kaon $ 
 &  $ \Lambda \Ds $\\
 2591.6  &  2806.15   &  2833.97   
 &  2947.54   &  2965.11   &  3069.11   
 &  3072.51   &  3084.18  \\\hline
 $ \Lambda \DsS $ &  $ \Sigmac \rho 
  $ &  $ \Lambdac \etap $ 
 &  $ \SigmacS \rho $ &  $ \Lambdac 
  \phi $ &  $ \Cascadac \kaonS $ 
 &  $ \Cascadacp \kaonS $ 
 &  $ \CascadacS \kaonS $\\
 3227.98  &  3229.05   &  3244.24   
 &  3293.46   &  3305.92   &  3361.11   
 &  3468.51   &  3538.01  \\
\hline
\end{tabular}
\begin{center}
{$I=1$, $J=1/2$} 
\end{center}
\begin{tabular}{@{\extracolsep{1mm}}cccccccccccccccccccc}
\hline
 $ \Lambdac \pion $ &  $ \Sigmac \pion 
  $ &  $ \Nucleon \D $ 
 &  $ \Nucleon \DS $ &  $ \Cascadac 
  \kaon $ &  $ \Sigmac \eta $ 
 &  $ \Lambdac \rho $ &  $ \Cascadacp 
  \kaon $ &  $ \Sigma \Ds $ 
 &  $ \Sigmac \rho $ &  $ \Sigmac \omega 
  $\\
 2424.5  &  2591.6   &  2806.15   
 &  2947.54   &  2965.12   &  3001.07   
 &  3061.95   &  3072.52   &  3161.64   
 &  3229.05   &  3236.21  \\\hline
 $ \Delta \DS $ &  $ \SigmacS \rho 
  $ &  $ \SigmacS \omega $ 
 &  $ \Sigma \DsS $ &  $ \Cascadac 
  \kaonS $ &  $ \Sigmac \etap $ 
 &  $ \Cascadacp \kaonS $ 
 &  $ \Sigmac \phi $ &  $ \SigmaS \DsS 
  $ &  $ \SigmacS \phi $ 
 &  $ \CascadacS \kaonS $\\
 3240.62  &  3293.46   &  3300.62   
 &  3305.45   &  3361.11   &  3411.34   
 &  3468.51   &  3473.01   &  3496.87   
 &  3537.42   &  3538.01  \\
\hline
\end{tabular}
\begin{center}
{$I=0$, $J=3/2$} 
\end{center}
\begin{tabular}{@{\extracolsep{1mm}}cccccccccccccccccccc}
\hline
 $ \SigmacS \pion $ &  $ \Nucleon \DS 
  $ &  $ \Lambdac \omega $ 
 &  $ \CascadacS \kaon $ 
 &  $ \Lambda \DsS $ &  $ \Sigmac \rho 
  $ &  $ \SigmacS \rho $ 
 &  $ \Lambdac \phi $ &  $ \Cascadac 
  \kaonS $ &  $ \Cascadacp \kaonS $ 
 &  $ \CascadacS \kaonS $\\
 2656.01  &  2947.54   &  3069.11   
 &  3142.01   &  3227.98   &  3229.05   
 &  3293.46   &  3305.92   &  3361.11   
 &  3468.51   &  3538.01  \\
\hline
\end{tabular}
\begin{center}
{$I=1$, $J=3/2$} 
\end{center}
\begin{tabular}{@{\extracolsep{1mm}}cccccccccccccccccccc}
\hline
 $ \SigmacS \pion $ &  $ \Nucleon \DS 
  $ &  $ \Lambdac \rho $ 
 &  $ \SigmacS \eta $ &  $ \Delta \D 
  $ &  $ \CascadacS \kaon $ 
 &  $ \Sigmac \rho 
  $ &  $ \Sigmac \omega $  &  $ \Delta \DS $
 &  $ \SigmacS \rho $\\
 2656.01  &  2947.54   &  3061.95
 &  3065.48   &  3099.23   &  3142.02   
  &  3229.05   &  3236.21  &  3240.62  
 &  3293.46  \\\hline
 $ \SigmacS \omega $ &  $ \Sigma \DsS 
  $ &  $ \SigmaS \Ds $ 
 &  $ \Cascadac \kaonS $ &  $ \Cascadacp 
  \kaonS $ &  $ \Sigmac \phi $ 
 &  $ \SigmacS \etap $ &  $ \SigmaS 
  \DsS $ &  $ \SigmacS \phi $ 
 &  $ \CascadacS \kaonS $\\
 3300.62  &  3305.45   &  3353.06   
 &  3361.11   &  3468.51   &  3473.01   
 &  3475.75   &  3496.87   &  3537.42
 &  3538.01  \\
\hline
\end{tabular}
\label{table1}
\end{table}
where below every channel we indicate its mass threshold, $M+m$, in
MeV units. Compared to \cite{GarciaRecio:2008dp}, we take $m_\Delta=1232$ 
MeV, instead of the pole mass.  As a consequence, resonances that
couple strongly to channels with $\Delta$ component might slightly
change its position and width, as the case of ($I=1$,$J=1/2$)
$\Sigma_c(2556)$ and ($I=1$,$J=3/2$) $\Sigma_c(2554)$, few MeV's above
the values in Ref. ~\cite{GarciaRecio:2008dp}.  We also take $m_{K^*}=892 \ {\rm
MeV}$. Besides, we also use experimental, when possible, or
theoretical estimates for the meson  decay constants.
The used values in this work can be found in Table~II of
Ref.~\cite{GarciaRecio:2008dp}. 

With the kernel of the meson-baryon interaction given in
Eq.~(\ref{eq:vsu8break}), we obtain the coupled $DN$ and $D^*N$
effective interaction in free space by solving the on-shell
Bethe-Salpeter equation
\cite{OR98,Meissner:1999vr,Oller:2000fj,Jido:2003cb,EJmeson,EJmeson2}
\begin{eqnarray}
T^{IJ}(\sqrt{s}) &=& \frac{1}{1-
V^{IJ}(\sqrt{s})\,G^{0\,(IJ)}(\sqrt{s})}\,V^{IJ}(\sqrt{s}) \ ,
 \label{eq:scat-eq}
\end{eqnarray}
in the coupled channel space.  Here $G^{0\,(IJ)}(\sqrt{s})$ is a
diagonal matrix consisting of loop functions. The free space loop
function for channel $a$ reads
\begin{eqnarray}
G^{0\,(IJ)}_{a}(\sqrt{s}) &=& {\rm i} 2 M_a \int \frac{d^4 q}{(2 \pi)^4}~
{D^0_{{\cal B}_a}(P-q)~D^0_{{\cal M}_a}(q)
}\ ,\\
D^0_{{\cal B}_a}(P-q)&=&((P-q)^2-M_a^2+i \varepsilon)^{-1}\ ,\\
D^0_{{\cal M}_a}(q)&=&(q^2-m_a^2+i \varepsilon)^{-1} \ ,
\label{eq:gprop}
\end{eqnarray}
where $s=P^2$, $D^0$ is a free hadron propagator, $a$ runs for the
allowed baryon-meson channels for the given $IJ$ sector, and $M_a$ and
$m_a$ are the masses of the baryon ${\cal B}_a$ and meson ${\cal M}_a$
in the channel $a$, respectively.  The previously defined loop
function is ultraviolet (UV) divergent. However, the difference
$G^0(\sqrt{s_1})-G^0(\sqrt{s_2})$ is finite for any finite values of $s_1$
and $s_2$, hence, the function can be regularized by setting a finite
value of $G^0$ at a given point. We choose
\begin{equation}
G^{0\,(IJ)}_{a}(\sqrt{s}=\mu_a^{IJ}) = 0\,,
\label{eq:subs}
\end{equation}
with index $a$ running in the coupled channel space, as done in
Refs.~\cite{GarciaRecio:2008dp,hofmann,hofmann2}. In those works, the subtraction point was
taken to be independent of $a$ and $J$ as
\begin{equation}
\left (\mu^{I}\right)^2=\alpha
\left(m_{\text{th}}^2+M^2_{\text{th}}\right)\ ,
\label{eq:sp}
\end{equation}
where $m_{\text{th}}$ and $M_{\text{th}}$ are the meson and baryon
masses of the hadronic channel with lowest mass threshold for a fixed
$I$ and arbitrary $J$. The value of $\alpha=0.9698$ was adjusted to
reproduce the position of the well established $\Lambda_c(2595)$
resonance with $IJ=(0,1/2)$.  Then, the same value will be used in all
other sectors. In this work we will follow the same prescription,
taking $(\mu^{I=0})^2 = \alpha({M_{\Sigma_c}}^2+m_\pi^2)$ and
$(\mu^{I=1})^2 = \alpha ({M_{\Lambda_c}}^2+m_\pi^2)$.

Hence, the renormalized (finite) loop function finally reads:
\begin{eqnarray}
G^0(\sqrt{s}) &=& {\rm i} 2 M \int \frac{d^4 q}{(2 \pi)^4}
\left( D^0_{\cal B}(P-q)~D^0_{\cal M}(q)
-D^0_{\cal B}(\bar{P}-q)~D^0_{\cal M}(q) \right) \ ,
\label{eq:g0sustr}
\end{eqnarray}
with $P$ and $\bar P$ defined such that $P^2=s,~ {\bar P}^2 =
(\mu^I)^2$, where, for simplicity, the obvious isospin $I$, spin $J$
and channel $a$ labels have been omitted.  This is the (standard) method
we use to  renormalize the free baryon-meson loop.

The properties of $D$ and $D^*$ mesons in nuclear matter are obtained
by incorporating the corresponding medium modifications in the
effective $DN$ and $D^*N$ interactions. One of the sources of density
dependence comes from the Pauli principle acting on the
nucleons. Another source is related to the change of the properties of
mesons and baryons in the intermediate states due to the interaction
with nucleons of the Fermi sea.

Those changes are implemented by using the in-medium hadron
propagators instead of the corresponding free ones.  Therefore, we
should define a consistent renormalization scheme, similar to that
adopted in the free case, for the loop function in a nuclear medium
with density $\rho$.

Let us first review what has been done before.  The in-medium loop
function, $G^\rho_\Lambda$, used in
Refs.~\cite{angels-mizutani,tolos-angels-mizutani}, depends on a
regularization cutoff $\Lambda$ that renders the UV divergence finite, 
and it is defined as:
\begin{eqnarray}
G^\rho_\Lambda(P) &=& {\rm i} 2 M \int_\Lambda \frac{d^4 q}{(2 \pi)^4}~
D^\rho_{\cal B}(P-q)~D^\rho_{\cal M}(q)  \ ,
\end{eqnarray}
where $D^\rho_{\cal B},~D^\rho_{\cal M}$ are the hadron propagators in
a medium with density $\rho$. In those
works~\cite{angels-mizutani,tolos-angels-mizutani}, the three--momentum
cutoff $\Lambda$ is fixed in such a way that the free ($\rho=0$)
results reproduce certain known experimental results (for instance,
the position of the resonance $\Lambda_c(2595)$ for the sector with
quantum numbers $IJSC = 0,1/2,0,1$).  This way of regularizing the 
UV loop function induces  medium corrections of the type:
\begin{eqnarray}
G^\rho_\Lambda(P) &=& G^0_\Lambda (\sqrt{s})+ \delta G^\rho_\Lambda(P), \nonumber \\
\delta G^\rho_\Lambda(P) &\equiv& G^\rho_\Lambda(P)-G^0_\Lambda(\sqrt{s})~=~{\rm i}
2 M \int_\Lambda \frac{d^4 q}{(2 \pi)^4} \left ( D^\rho_{\cal
B}(P-q)~D^\rho_{\cal M}(q)-D^0_{\cal B}(P-q)~D^0_{\cal M}(q) \right ) \  .
\label{eq:deltaG}
\end{eqnarray}
In this work, we do want to avoid finite-cutoff effects.  So we define
the in-medium loop function as the free one $G^0$,  given in 
Eq.~(\ref{eq:g0sustr}) and defined  as in
Refs.~\cite{GarciaRecio:2008dp,hofmann,hofmann2,Garcia-Recio:2005hy,
QNP06,
Toki:2007ab} without having to introduce any cutoff, plus a term that
accounts for  the same kind of medium effects as those
displayed in Eq.~(\ref{eq:deltaG}),  but taking $\Lambda$ large enough (that is
$\Lambda \to\infty$). Hence, we will use:
\begin{eqnarray}
%\nonumber
G^\rho(P) &=& G^0(\sqrt{s}) + \delta G^\rho(P) \ , \nonumber \\
\label{eq:defGrhob}
\delta G^\rho(P) &=& \lim_{\Lambda\to\infty}\delta G^\rho_\Lambda(P)
~\equiv~{\rm i}
2 M \int \frac{d^4 q}{(2 \pi)^4} \left ( D^\rho_{\cal
B}(P-q)~D^\rho_{\cal M}(q)-D^0_{\cal B}(P-q)~D^0_{\cal M}(q) \right ) .
\end{eqnarray} 
The UV finite $\delta G^\rho$ correction contains all the nuclear
medium effects affecting the loop, and it is 
independent of the selected subtracting point used to regularizate
it. Then, a cutoff is not needed for the calculation.  The
defined loop function at finite density, $G^\rho$, can be rewritten as
\begin{eqnarray}
G^\rho(P) &=& {\rm i} 2 M \int \frac{d^4 q}{(2 \pi)^4}~
\left( {D^\rho_{{\cal B}}(P-q)~D^\rho_{{\cal M}}(q)
}-D^0_{\cal B}(\bar P-q)~D^0_{\cal M}(q) \right )\ ,
\label{eq:gpropsustr}
\end{eqnarray}
where the integrand is the difference between two terms.  The first
one corresponds to the baryon and meson propagators calculated at
density $\rho$ and total momentum $P$. The second term depends neither
on the density nor on $P$, and it is constructed out of propagators
evaluated at $\rho=0$ and with fixed total momentum $\bar P$,  such that
${\bar P}^2 = (\mu^I)^2$.  This shows that our prescription of
Eq.~(\ref{eq:defGrhob}) for $G^\rho$ amounts to assume that the 
employed subtraction to make the UV divergent function finite is
 independent of the nuclear density.

For practical numerical purposes we will calculate $G^\rho$ as specified in
Eq.~(\ref{eq:defGrhob}), where the free part is
analytical, regularized with a subtracting constant and well known,
and the medium modification part $\delta G^\rho$ is numerically
evaluated in the same way as done in previous works for $G^\rho_\Lambda$, but 
taking into account that it has a subtracting part $G^0$ providing a  
$\delta G^\rho$ that is UV finite. 

For the $DN$ and $D^*N$ channels, we consider Pauli blocking effects
on the nucleons together with self-energy insertions of the $D$ and
$D^*$ mesons. The self-energy is obtained self-consistently from the
in-medium $DN$ and $D^*N$ effective interactions,
${T^\rho}_{D(D^*)N}$, as we will show in the following. The
corresponding in-medium single-particle propagators are given by:
\bea
D^\rho_N(p)~&=& {1\over 2 E_N(\vec{p}\,)}
\left (\frac{1-n(\vec{p}\,)}{p^0 -E_N(\vec{p}\,) + i \varepsilon}+
\frac{n(\vec{p}\,)}{p^0 - E_N(\vec{p}\,)-i \varepsilon} +
\frac{1}{-p^0 -E_N(\vec{p}\,) + i \varepsilon}\right ) \nonumber \\ 
&=& D^0_N(p)~+~2\pi i~n(\vec{p}\,)~\frac{\delta\left (p^0-E_N({\vec p}\,)\right )}{2E_N(\vec{p}\,)}\ ,\\
D^\rho_{D (D^*)}(q)&=&\left ((q^0)^2 -\omega({\vec q})^2-\Pi_{D(D^*)}(q) \right )^{-1}\,=\, 
\int_{0}^{\infty} \, d\omega \, \left(
\frac{S_{D(D^*)}(\omega,\vec{q}\,)}{q^0-\omega +i
\varepsilon}-\frac{S_{\bar{D}(\bar{D}^*)}(\omega,\vec{q}\,)}{q^0+\omega-i
\varepsilon} \right) ,~~
\label{eq:Drho}
\eea
where $E_N(\vec{p}\,)=\sqrt{{\vec{p}\,}^2+M_N^2}$~,~~ $ \omega({\vec
q}\,)=\sqrt{{\vec{q}\,}^2+m_{D(D^*)}^2}$~,
$\Pi_{D(D^*)}(q^0,\vec{q}\,)$ is the $D(D^*)$ meson self-energy and
$S_{D(D^*)}$ the corresponding meson spectral function. In a very good
approximation the spectral function for $\bar D$  can be approximated
by the free-space one, viz. by a delta function, because for that case
$C=-1$, there are no low-lying baryon resonances. Finally,
$n(\vec{p}\,)$ is the Fermi gas nucleon momentum distribution, given
by the step function $n(\vec{p}\,) = H(k_F-|\vec{p}\,|)$, with
$k_F=(3\pi^2\rho/2))^{1/3}$.

Using Eq.~(\ref{eq:defGrhob}) and performing the energy integral
over $q^0$, the $DN$ and $D^*N$ loop functions read
\begin{eqnarray}
{G^\rho}_{D(D^*)N}(P)=
 G^0_{D(D^*)N}(\sqrt{s})+
\int \frac{d^3 q}{(2 \pi)^3} \,
\frac{ M_N }{ E_N(\vec{p}\,)} \,
\left [ 
\frac{-n(\vec{p}\,)}{(P^0 - E_N(\vec{p}\,))^2-\omega(\vec{q}\,)^2+i\varepsilon} 
\, +\right.\phantom{largoooooooooo} &&
\label{eq:Glarga}\\
\left.
(1-n(\vec{p}\,))
\left. \left (
\frac{-1/(2 \omega({\vec q}\,))}
{P^0 -E_N(\vec{p}\,)-\omega(\vec{q}\,)+i \varepsilon}
+
\int_{0}^{\infty} \,
 d\omega \,
\frac{S_{D(D^*)}(\omega,\vec{q}\,)}{P^0 -E_N(\vec{p}\,)-\omega+i\varepsilon}
\, \right ) \right ]  \right| _{{\vec p}={\vec P}-{\vec q}}\ ,&& 
\nonumber 
\end{eqnarray}
where the first term of the integral, proportional to $-n(\vec{p})$, is 
just what we call Pauli correction and accounts for the 
case where the Pauli blocking on the nucleon is considered and 
the meson in-medium selfenergy is neglected.  The  
second term, proportional to $(1-n(\vec{p}))$, is exactly zero if 
the meson spectral functions $S_{D(D^*)}$ were taken to be the free one, 
$S^{\rm free}_{D(D^*)}(\omega,\vec{q})=
\delta(\omega- \omega(\vec{q}))/(2 \omega)$.
Then, it accounts for the contribution of the in-medium meson modification to 
the loop function.  Note that, compared to the 
Refs.~\cite{angels-mizutani,tolos-angels-mizutani}, we also include
 the antiparticle contributions in the propagators.  

As for $D\Delta$ and $D^*\Delta$ channels, we include the self-energy
of the $D$ and $D^*$ mesons. Then, the equivalent of
Eq.~(\ref{eq:Glarga}) for those channels read
\begin{eqnarray}
&&{G^\rho}_{D(D^*)\Delta}(P)=
 G^0_{D(D^*)\Delta}(\sqrt{s}) \label{eq:propDD}\\ 
&+&\int \frac{d^3 q}{(2 \pi)^3} \,
\frac{ M_\Delta }{ E_\Delta(\vec{p}\,)} \,
\left. \left (
\frac{-1/(2 \omega({\vec q}\,))}
{P^0 -E_\Delta(\vec{p}\,)-\omega(\vec{q}\,)+i \varepsilon}
+
\int_{0}^{\infty} \,
 d\omega \,
\frac{S_{D(D^*)}(\omega,\vec{q}\,)}{P^0 -E_\Delta(\vec{p}\,)-\omega+i\varepsilon}
\, \right )  \right| _{{\vec p}={\vec P}-{\vec q}}\ .
\nonumber
\end{eqnarray}
For the other channels that couple to $DN$ and $D^*N$ (see
Table~\ref{table1}), we refrain from including any medium modifications
in the loop function and, therefore, we use the free-space one given
in Eq.~(\ref{eq:g0sustr}). This is due to the lack of knowledge on how
the properties of some mesons, such as $\rho$ or $\omega$, change in
the medium. Only pions in nuclear matter have been intensively studied
\cite{Nieves:1993ev,Nieves:1991ye}, but, as indicated in
Ref.~\cite{GarciaRecio:2008dp} and discussed in the next section, the coupling to
intermediate states with pions is of minor importance for the
dynamical generation of the baryon resonances in the $S=0$ and $C=1$
sector that governs the $DN$ and $D^*N$ dynamics in nuclear matter.

We can now solve the on-shell Bethe-Salpeter equation in nuclear matter for 
the in-medium amplitudes
\begin{eqnarray}
{T^\rho}^{(IJ)}(P) &=& \frac{1}{1-
V^{IJ}(\sqrt{s})\,{G^\rho}^{(IJ)}(P)}\,V^{IJ}(\sqrt{s}) \ .
 \label{eq:scat-rho}
\end{eqnarray}
The in-medium $D$ and $D^*$ self-energies are finally obtained by
integrating ${T^\rho}_{D(D^*)N}$ over the nucleon Fermi sea, 
\begin{eqnarray}
\Pi_D(q^0,\vec{q}\,)&=\int \frac{d^3p}{(2\pi)^3} \, n(\vec{p}\,)\,
&\left [~{T^\rho}^{(I=0,J=1/2)}_{DN}(P^0,\vec{P})+3 {T^\rho}^{(I=1,
J=1/2)}_{DN}(P^0,\vec{P})\right ] \ , \label{eq:pid}\\
\Pi_{D^*}(q^0,\vec{q}\,)&=\int \frac{d^3p}{(2\pi)^3} \, n(\vec{p}\,) \,
\, &\left [~ \frac{1}{3} \, {T^\rho}^{(I=0,J=1/2)}_{D^*N}(P^0,\vec{P})+
{T^\rho}^{(I=1,J=1/2)}_{D^*N}(P^0,\vec{P}) \right .\nonumber\\ &&+ \frac{2}{3} \,
\left . {T^\rho}^{(I=0,J=3/2)}_{D^*N}(P^0,\vec{P})+ 2 \,
{T^\rho}^{(I=1,J=3/2)}_{D^*N}(P^0,\vec{P})\right ] \  ,
\label{eq:pids}
\end{eqnarray}
where $P^0=q^0+E_N(\vec{p}\,)$ and $\vec{P}=\vec{q}+\vec{p}$ are the
total energy and momentum of the $DN$ ($D^*N$) pair in the nuclear
matter rest frame and the values $(q^0,\vec{q}\,)$ stand for the
energy and momentum of the $D$ and $D^*$ meson also in this frame. The
$\Pi_{D(D^*)}(q^0,\vec{q}\,)$ has to be determine self-consistently
since it is obtained from the in-medium amplitude $
{T}^\rho_{D(D^*)N}$ which contains the $D(D^*)N$ loop function
${G}^\rho_{D(D^*)N}$, and this last quantity itself is a function of
$\Pi_{D(D^*)}(q^0, \vec q \,)$. From this we obtain the corresponding
spectral function to complete the integral for the loop function
${G^\rho}_{D(D^*) (N,\Delta)}(P^0, \vec{P} \,)$ as given 
in Eqs.~(\ref{eq:Glarga},\ref{eq:propDD}).

\section{Results}

\begin{figure}[t]
\begin{center}
\includegraphics[width=0.8\textwidth]{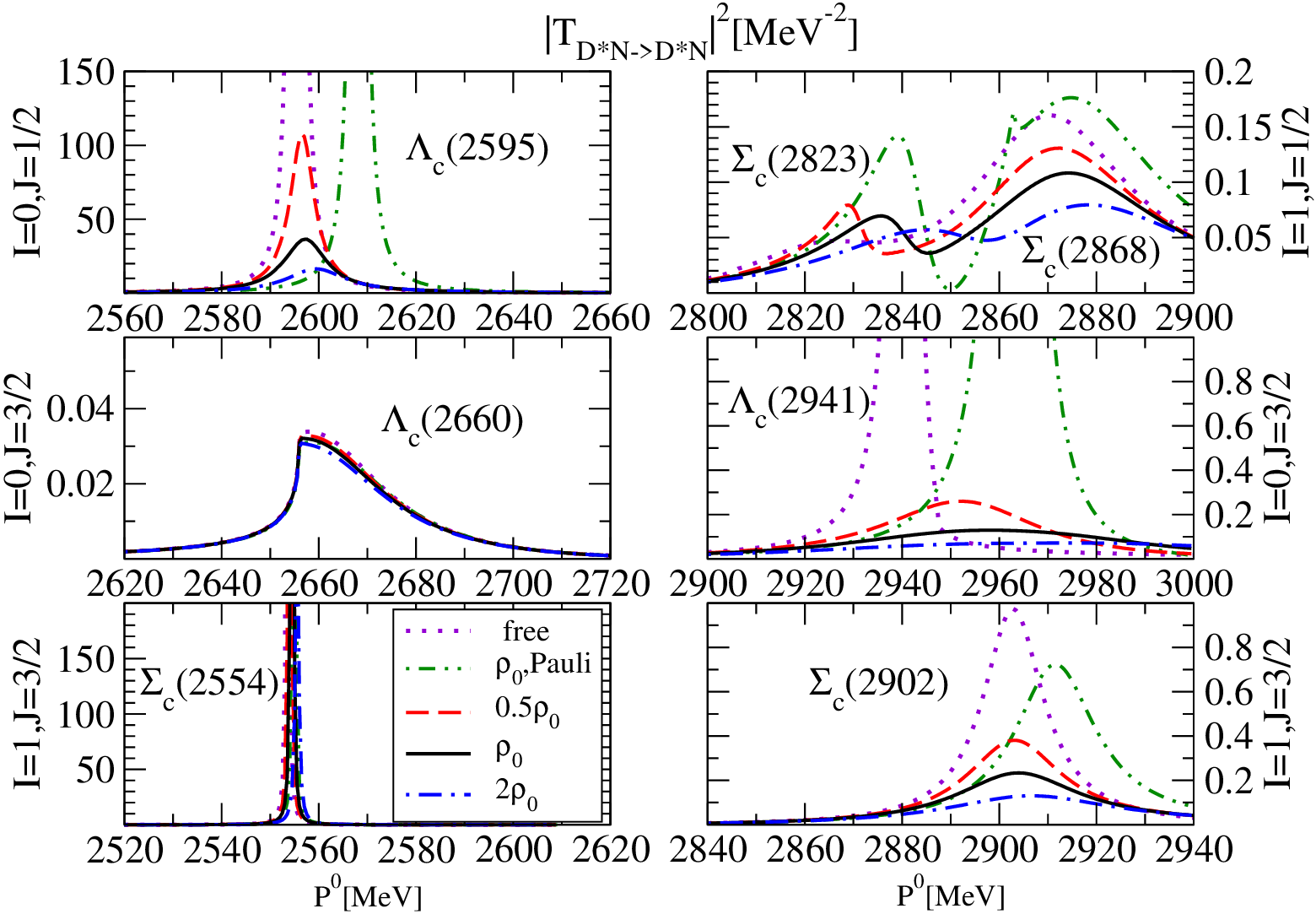}
\caption{Squared of the $D^*N \rightarrow D^* N$ amplitude for different
partial waves as function of the center-of-mass energy $P^0$ for fixed
total momentum $|\vec{P}|=0$. Several resonances are shown: ($I=0$,
$J=1/2$) $\Lambda_c(2595)$, ($I=1$,$J=1/2$) $\Sigma_c(2823)$ and
$\Sigma_c(2868)$, ($I=0$, $J=3/2$)  $\Lambda_c(2660)$, ($I=0$, $J=3/2$)
$\Lambda_c(2941)$, ($I=1$, $J=3/2$) $\Sigma_c(2554)$ and ($I=1$,
$J=3/2$) $\Sigma_c(2902)$.}
\label{fig:reso}
\end{center}
\end{figure}

We start this section by displaying in Fig.~\ref{fig:reso} the squared
amplitude of the $D^*N \rightarrow D^* N$ transition for different
partial waves as a function of the center-of-mass energy $P^0$ for a
total momentum $|\vec{P}|=0$. In particular, we show certain partial
waves and energy regimes where we can find seven resonances predicted
by the SU(8) model \cite{GarciaRecio:2008dp} that have or can have experimental
confirmation \cite{Yao}, i.e., ($I=0$, $J=1/2$) $\Lambda_c(2595)$,
($I=1$,$J=1/2$) $\Sigma_c(2823)$ and $\Sigma_c(2868)$, ($I=0$,
$J=3/2$) $\Lambda_c(2660)$,($I=0$, $J=3/2$) $\Lambda_c(2941)$, ($I=1$,
$J=3/2$) $\Sigma_c(2554)$ and ($I=1$, $J=3/2$) $\Sigma_c(2902)$
resonances. All of them couple to the $D^*N$ despite not being the
dominant one for $\Lambda_c(2660)$, $\Sigma_c(2823)$ and
$\Sigma_c(2554)$, as discussed in Ref.~\cite{GarciaRecio:2008dp}.  However, we choose to
display these amplitudes for different nuclear densities, since  they
 the determine the $D^*$ self-energy, as follows from
 Eq.~(\ref{eq:pids}). We analyze three different cases: (i) solution of the
on-shell Bethe-Salpeter equation in free space (dotted lines), which
was already studied in Ref.~\cite{GarciaRecio:2008dp}, (ii) in-medium calculation
of the on-shell Bethe-Salpeter including Pauli blocking on the nucleon
intermediate states at normal nuclear matter density $\rho_0=0.17
~ \rm{fm}^{-3}$ (dashed lines), (iii) in-medium solution which
incorporates Pauli blocking effects and the $D$ and $D^*$
self-energies in a self-consistent manner for three densities, ranging
from $0.5$ to $2$ $\rho_0$ (solid lines).

\begin{figure}[t]
\begin{center}
\includegraphics[width=0.5\textwidth,]{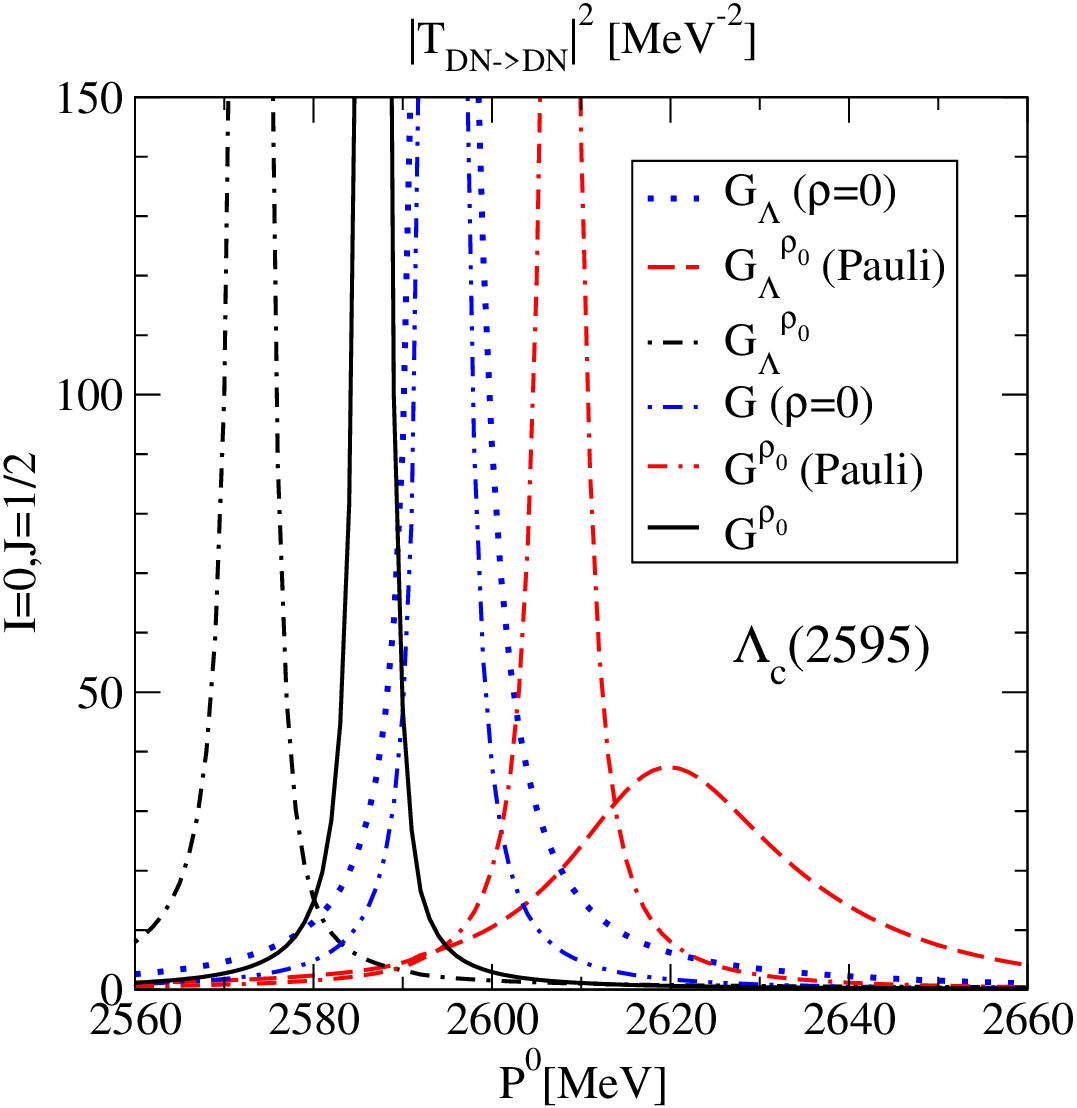}
\caption{Squared of the $DN \rightarrow D N$ amplitude in ($I=0$, $J=1/2$)
$\Lambda_c(2595)$ channel as function of the center-of-mass energy
  $P^0$ with $|\vec{P}\,|=0$. We show results obtained with 
the SU(4) TVME model with a three-momentum cutoff ($\Lambda=787 \, {\rm MeV/c}$), as in
Ref.~\cite{angels-mizutani}, and for
comparison, we also show results using the renormalization scheme
assumed in this work ($\alpha=0.895$) [Eqs.~(\ref{eq:defGrhob},\ref{eq:gpropsustr})]. We examined 
three different cases: free amplitude, calculation including
Pauli blocking effects at $\rho=\rho_0$ and in-medium self-consistent
solution at $\rho_0$.}
\label{fig:reso_su4}
\end{center}
\end{figure}

The $\Lambda_c(2595)$ resonance is predominantly a $D^*N$ bound state
in contrast to SU(4) TVME model, where it emerged as a $DN$ quasi-bound
state
\cite{kolo,hofmann,hofmann2,lutz,angels-mizutani,tolos-angels-mizutani}. Pauli
blocking effects on the intermediate nucleon states move the resonance
to higher energies, as already found in previous in-medium models
\cite{tolos-schaffne-mishra,angels-mizutani,lutz}, due to the
restriction of available phase space in the unitarization
procedure. The shift in mass is of the order of 12 MeV, while for the
SU(4) TVME model of Ref.~\cite{angels-mizutani} the shift is around a
factor of two larger. This change in the energy shift provided by
Pauli blocking can be attributed either to the different SU(4) and
SU(8) kernels, or to the renormalization scheme employed to make
finite the loop function or the combination of both effects. In order
to disentangle between them, we study in Fig.~\ref{fig:reso_su4} the
$\Lambda_c(2595)$ resonance in the $DN \rightarrow DN$ transition by using the SU(4) TVME model (with $\Sigma_{DN}$=0)  of
Ref.~\cite{angels-mizutani} for cases (i) to (iii) above.  We compare
results obtained by using the cutoff
regularization~\cite{angels-mizutani} and our new renormalization
scheme [Eqs.~(\ref{eq:defGrhob},\ref{eq:gpropsustr})]\footnote{We fix now $\alpha=0.895$
to obtain the correct position of $\Lambda_c(2595)$}.  A similar mass
shift for the $\Lambda_c(2595)$ is observed in both SU(4) and SU(8)
models when Pauli blocking effects are included for the new
renormalization scheme. Therefore, we conclude that the different
renormalization of the loop function is the main source of
discrepancy, i.e., the in-medium solution depends strongly on the
correct treatment of the $\sqrt{s}$ dependence of the loop function. The treatment proposed here clearly
improves over those based on the use of a cutoff. Thus, the resonances which are
far offshell from their dominant channel will be heavily affected, as
in the case of $\Lambda_c(2595)$. The self-consistent procedure moves
the resonance closer to the free position because the repulsive effect
of the Pauli blocking is tamed by the inclusion of the $D$ and $D^*$
self-energies. In this case, the difference between SU(8) and SU(4)
models is not only due to the renormalization scheme but also to the
inclusion of the $D^*$ self-energy, which compensates the attraction
felt by the $D$ mesons, as we will see in the next figures.
  
The ($I=1$,$J=1/2$) $\Sigma_c(2823)$ and $\Sigma_c(2868)$ are shown in
 the top right panel of Fig.~\ref{fig:reso}. Although no experimental
evidence of those resonances is available yet, they lie very close to
the $DN$ threshold and, therefore, changes in the nuclear medium will
have an important effect on the $D$ self-energy, which is a matter of
interest in this work. In fact, those resonant states are
significantly modified in the medium, since both resonances couple
significantly to $DN$ and $D^*N$ systems as well as $D^*\Delta$ and
any medium modification in those systems alters their behavior.
  
The next resonance predicted by the SU(8) model \cite{GarciaRecio:2008dp}, the
($I=0$,$J=3/2$) $\Lambda_c(2660)$, might be identified with the
experimental $\Lambda_c(2625)$, which is the charm counterpart of the
$\Lambda(1520)$. This state couples strongly to the $\Sigma^*_c \pi$
channel and more weakly to the $D^*N$ pair. Pauli blocking and
self-consistency have smaller effects than in the case of the
$\Lambda_c(2595)$ resonance. This is because, while the
$\Lambda_c(2595)$ resonance varies in the medium due to the changes
affecting its two dominant channels, $D^*N$ and $DN$, the
$\Lambda_c(2660)$ is only modified via the secondary $D^*N$
channel. We do not include any medium modifications affecting to its
dominant channel, $\pi \Sigma^*_c$. Though the pion self-energy
\cite{Nieves:1993ev,Nieves:1991ye} may induce some changes in
this resonance, the expected effect of those modifications in the $D$
and $D^*$ self-energies are minor. On one hand, this partial wave will
only have a direct contribution to the $D^*$ self-energy, and  thus the $D$
self-energy will be affected indirectly via the simultaneous
self-consistent calculation of the $D$ and $D^*$ self-energies. On the
other hand, the effect of this resonance in the $D^*$ self-energy is
marginal because it only reflects in the low-energy tail, far from the
quasiparticle peak. Then, we refrain to introduce any medium changes
for this channel.

The ($I=0$,$J=3/2$) $\Lambda_c(2941)$ resonance might be a candidate
for the experimental $\Lambda_c(2940)$, which $J^P$ is unknown
\cite{Yao}. This correspondence is made under the assumption that our
model needs an additional implementation of $p$-wave interactions in
order to explain the decay into $D^0p$ pairs reported in
Ref.~\cite{2940-Aubert:2006sp}, which is also hinted by the dominant
coupling to $D^*N$. This strong coupling changes its properties
significantly when medium modifications are implemented. Moreover, the
fact that the $\Lambda_c(2941)$ lies so close to the $D^*N$ threshold
will have important consequences on the $D^*$ self-energy and, hence,
on the spectral function, as we will see in the following.
 
The ($I=1$,$J=3/2$) $\Sigma_c(2554)$ resonance has not yet a experimental
confirmation. However, similarly to the $\Lambda_c(2660)$, this
resonance might be the counterpart in the charm sector of the
$\Sigma(1670)$. It couples strongly to $D\Delta$ and $D^*\Delta$
channels, which correspond to the $\bar K \Delta$ in the strange
sector. Pauli blocking effects on the nucleons are relatively weak
because the coupling to the $D^*N$ channel is approximately half the
coupling to the two dominant channels. Changes due to the
selfconsistent procedure are comparable to the case of the
$\Lambda_c(2660)$.

The $\Sigma_c(2902)$ resonance in the ($I=1, J=3/2$) sector can be a
candidate for the $\Sigma_c(2800)$ resonance, by varying slightly the
renormalization scale and if this resonance could be also seen in $\pi
\pi \Lambda_c$ states \cite{GarciaRecio:2008dp}. This is in contrast with SU(4)
TVME models, which predict it in the $I=1, J=1/2$ channel
\cite{hofmann,hofmann2,lutz,angels-mizutani}.  With regard to medium
effects, those are comparable to the $\Lambda_c(2595)$ case. The
dominant channel for the generation of this resonance is
$D^*N$. Moreover, this resonance lies 50 MeV below the $D^*N$
threshold. Therefore, modifications due to Pauli blocking and
self-consistency are expected to be more important than for
$\Lambda_c(2660)$ and $\Sigma_c(2554)$, and turn out to be comparable
to the changes in $\Lambda_c(2595)$.

\begin{figure}[t]
\begin{center}
\includegraphics[width=0.8\textwidth]{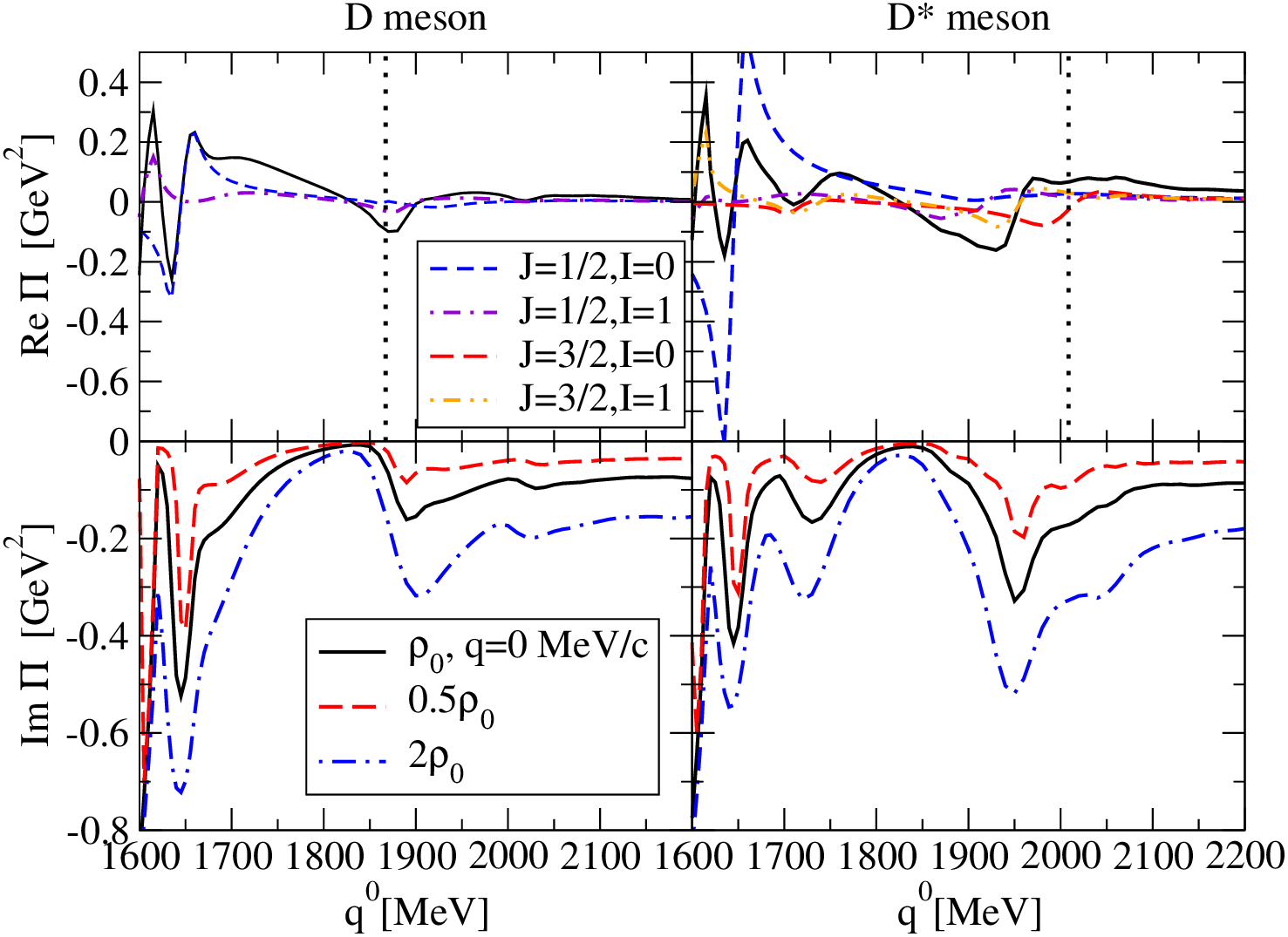}
\caption{Real and imaginary parts of the $D$ and $D^*$ self-energies
as functions of the meson energy $q^0$, including the decomposition
in partial waves (first row), and for different densities at $q=0 \, {\rm MeV/c}$
(second row). The positions of the $D$ and $D^*$ meson masses
  are also shown as vertical lines for reference.}
\label{fig:auto}
\end{center}
\end{figure}

We show next in the left and right panels of Fig.~\ref{fig:auto} the
$D$ and $D^*$ self-energies, respectively, as  functions
of the meson energy $q^0$. The $D$ and $D^*$ self-energies result from
the integration over the $DN$ and $D^*N$ amplitudes, respectively,
after self-consistency is reached simultaneously. In the upper panels
we display the real part of the self-energies for $\rho_0$ at meson
zero momentum (solid lines) together with the partial wave
decomposition (dashed and dash-dotted lines). The partial waves weighted by
the corresponding factors in Eqs.~(\ref{eq:pid}) and~(\ref{eq:pids}) are
summed up in order to obtain the total self-energies. In the lower
panels we show the imaginary part of the self-energies for densities
ranging from 0.5 $\rho_0$ to 2 $\rho_0$ and $q=0 \ {\rm MeV/c}$ (solid
lines). The dotted vertical lines in the upper panels indicate
the free $D$ and $D^*$ meson masses. 

With regard to the $D$ meson self-energy, we observe that in the $I=0,
J=1/2$ partial wave the contribution of  
the $\Lambda_c(2595)$ resonance clearly appears
for energies of the $D$ meson around $1650 \ {\rm MeV}$, while the
resonant state $\Sigma_c(2556)$ governs the $I=1,J=1/2$ partial wave around
$q^0=1615 \ {\rm MeV}$. This  state couples mostly to $D^*\Delta$,
mixed with $DN$ and $D_s^*\Sigma^*$, and it is absent in the SU(4)
models \cite{kolo,hofmann,hofmann2,lutz,angels-mizutani}, which do not
include channels with a vector meson and a $3/2^+$ baryon. Close to
the $DN$ threshold ($\sqrt{s}=2806 \ {\rm MeV}$ in free space), the
$I=1, J=1/2$ is the dominant partial wave. This is a consequence of
the fact that this threshold lies very close to a resonant state in
the sector $I=1, J=1/2$ of $2823 \ {\rm MeV}$ with a width of $\Gamma=35 \ {\rm
MeV}$. This resonance is affected by Pauli blocking and
self-consistency, i.e, by the nuclear medium, as it couples strongly to
states with $D$, $D^*$ and nucleon content. Closer to that structure,
we also found a very narrow state in $I=0,J=1/2$ with a mass of $2821\
{\rm MeV}$. This resonant state is less modified in the medium, since it
couples marginally to the $DN$ channel. Due to its narrow width even
in nuclear matter, the main contribution to the $D$ meson self-energy
close to the $DN$ threshold comes from the $J=1/2$ $\Sigma_c(2823)$
resonance but modified by the near resonance $J=1/2$
$\Sigma_c(2868)$. Those resonances lie above the $DN$ threshold and,
hence, have an attractive effect at the $DN$ threshold.

The imaginary part of the $D$ meson self-energy, in absolute value,
 grows with increasing density because of an enhancement of collision
 and absorption processes. The change
 in density can be seen more easily in the spectral
 function, as we will show below in Fig.~\ref{fig:spec}. Compared to
 previous results in nuclear matter
 \cite{tolos-schaffne-mishra,lutz,angels-mizutani,tolos-angels-mizutani,tolos-schaffne-stoecker},
 the density  dependence of the $D$ meson self-energy are
 qualitatively similar. However, in the SU(8) model, we have a 
 richer spectrum of resonant states which is reflected in the
 self-energy. While the $\Lambda_c(2595)N^{-1}$ and
 $\Sigma_c(2800)N^{-1}$ determine the $D$ meson self-energy in SU(4)
 models \cite{lutz,angels-mizutani}, those contributions together
 with few other resonant-hole states around $q^0=1860-2060 \ {\rm MeV}$,
 such as $\Sigma_c(2823)N^{-1}$ and $\Sigma_c(2868)N^{-1}$, clearly 
 manifest  also in the $D$ self-energy using the SU(8) interaction
 \cite{GarciaRecio:2008dp}.

A novelty of the SU(8) model is that it allows to simultaneously
obtain the $D$ and $D^*$ self-energies. The $D^*$ self-energy comes
from the contribution not only from the $J=1/2$ partial waves but also
from the $J=3/2$ ones of the $D^*N$ amplitude. As expected in the
$J=1/2$ sector, we find the $\Lambda_c(2595)N^{-1}$ and
$\Sigma_c(2556)N^{-1}$ components for $q^0=1650 \ {\rm MeV}$ and
$q^0=1615 \ {\rm MeV}$, respectively. For higher energies around
$q^0=1880-1930$ we see the $\Sigma_c(2823)N^{-1} $ and
$\Sigma_c(2868)N^{-1}$ contributions. In the $J=3/2$ sector, we find
$\Sigma_c(2554)N^{-1}$ for $q^0=1610 \ {\rm MeV}$. Close to the $D^*N$
threshold ($\sqrt{s}=2947 \ {\rm MeV}$ in free space), around
$q^0=1960 \ {\rm MeV}$, the $\Sigma_c(2902)N^{-1}$ excitation becomes
dominant. This resonant-hole state mixes with $\Lambda_c(2941)N^{-1}$.
The combination of both $J=3/2$ resonances becomes the dominant
contribution close to $D^*N$ threshold and has a repulsive effect in
the $D^*$ self-energy, as observed in the upper right panel. Density effects have a similar outcome as in the case of the $D$
self-energy. Those effects are better visualized with the spectral
function.

\begin{figure}[t]
\begin{center}
\includegraphics[width=0.8\textwidth]{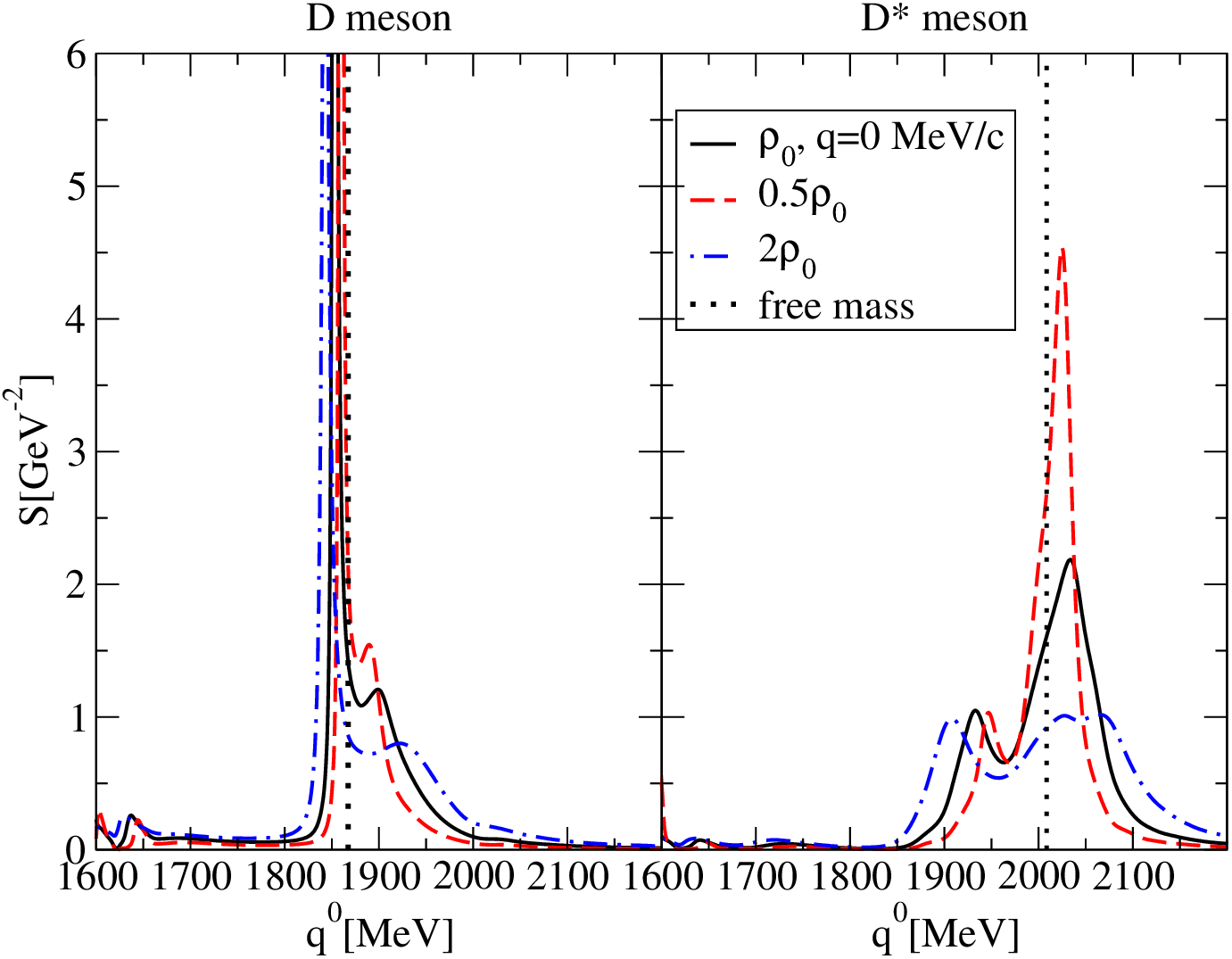}
\caption{$D$ and $D^*$ spectral functions as function of the meson
  energy $q^0$ for different densities at $q=0 \, {\rm MeV/c}$. The positions of the $D$ and $D^*$ meson free masses
  are also shown for reference (dotted vertical lines).}
\label{fig:spec}
\end{center}
\end{figure}

The spectral functions for $D$ and $D^*$ mesons as function of the
meson energy $q^0$ are displayed in the left and right panels of
Fig.~\ref{fig:spec}, respectively. The solid lines correspond to the
spectral functions at zero momentum from 0.5 $\rho_0$ to 2 $\rho_0$
for $|\vec{q}|=0 \ {\rm MeV/c}$. A dotted
vertical line indicating the free $D$ and $D^*$ meson masses is also drawn
for reference.

The quasiparticle peak of the spectral function, which is defined as
\begin{equation}
\omega_{qp}(\vec{q}\,)^2=\vec{q}\,^2+m^2+ {\rm Re}\Pi(\omega_{qp}(\vec{q}\,),\vec{q}\,) \ ,
\end{equation} 
moves to lower energies with respect to the free mass position for the
$D$ meson as density increases. As mentioned in Fig.~\ref{fig:auto},
the presence of the $J=1/2$ $\Sigma_c(2823)$ and $\Sigma_c(2868)$
resonances above threshold have an attractive effect at the $DN$
threshold. In fact, those resonances can be clearly seen on the
right-hand side of the quasiparticle peak. In the low-energy tail of
the $D$ spectral function, for energies around $1600-1650 \ {\rm
MeV}$, we observe the $\Sigma_c(2556)N^{-1}$ and
$\Lambda_c(2595)N^{-1}$ excitations. Moreover, other wider resonances
are generated in the SU(8) model \cite{GarciaRecio:2008dp} and they combine to give
the total $D$ meson spectral function. In the SU(4) TVME models
\cite{lutz,angels-mizutani,tolos-angels-mizutani}, the $J=1/2$
$\Sigma_c(2800)N^{-1}$ fully mixes with the quasiparticle peak while
the $\Lambda_c(2595)N^{-1}$ appears at the same energies as in our SU(8)
model, as expected.

The quasiparticle peak of the $D^*$ spectral function moves to higher
energies with density and fully mixes with the sub-threshold $J=3/2$
$\Lambda_c(2941)$ resonance. In the left-hand side of the peak we
observe the mixing of $J=1/2$ $\Sigma_c(2868)N^{-1}$ and $J=3/2$
$\Sigma_c(2902)N^{-1}$ excitations.  Other dynamically-generated
particle-hole states appear for higher and lower energies, such as
$J=3/2$ $\Sigma_c(2554)N^{-1}$.

Density effects result in a broadening of the spectral functions as
the collisional and absorption processes increase together with a
dilution of the resonant-hole states. This outcome is qualitatively similar to previous
models for the $D$ meson spectral function
\cite{lutz,angels-mizutani,tolos-angels-mizutani,tolos-schaffne-mishra,tolos-schaffne-stoecker}.

As already mentioned in one of the above references \cite{tolos-angels-mizutani}, the low-energy tail
of the $D$ meson spectral function due to resonant-hole states
($\tilde Y_c N^{-1}$) might help to understand the $J/\Psi$ suppression in
an hadronic scenario. However, it is unlikely that this lower tail
extends with sufficient strength as far as the $J/\Psi$ threshold to
explain $J/\Psi$ suppression only via the $D \bar D$ decay. A more
plausible hadronic contribution for the $J/\Psi$ suppression is the
reduction of its supply from the excited charmonia, $\chi_{c\ell}(1P)$
or $\Psi'$, which may find in the medium other competitive decay
channels \cite{tolos-angels-mizutani}. Such a more broad scenario for the $J/\Psi$ suppression has
been pictured recently in thermal models \cite{munzinger}. On the
other hand, the spectral function for the $D$ and $D^*$ mesons will
influence the behavior of dynamically-generated hidden and open charm
scalar resonances in nuclear matter, as already pointed out in
Ref.~\cite{raquel}.

 \begin{table}[tb]
    \centering
    \caption{$DN$ and $D^* N$ scattering lengths ({\rm fm})}
   \begin{tabular}{l |  c |  c  }
     & $DN$ & $D^* N$ \\
\hline
    ~~$J=1/2$ $I=0$~~ & ~~0.001 + i 0.002~~  &  ~~$-$0.44 + i 0.19~~   \\
    ~~(Born approx.)~~  & ~~( 0.59 + i 0 )~~  &  ~~( 1.82 + i 0 )~~   \\
\hline
    ~~$J=1/2$  $I=1$~~ & ~~0.33 + i 0.05 ~~& ~~$-$0.36  + i 0.18  ~~ \\
    ~~(Born approx.)~~ & ~~( 0.20 + i 0 ) ~~& ~~ ( 0.07 + i 0 )  ~~ \\
\hline
    ~~$J=3/2$ $I=0$~~  & &~~$-$1.93 + i 0.19 ~~  \\
    ~~(Born approx.)~~  & &~~( 0 + i 0 ) ~~  \\
\hline
    ~~$J=3/2$  $I=1$~~ &   & ~~$-$0.57 + i 0.15 ~~\\
    ~~(Born approx.)~~ &   & ~~( 0.27 + i 0 ) ~~ \\
    \end{tabular}
    \label{table2}
\end{table}

Finally we  study the properties of $D$ and $D^*$ mesons close to
the $DN$ and $D^*N$ threshold. We first present in Table~\ref{table2}
results for the $DN$ and $D^*N$ effective interactions in free
space. In particular, we give the $J=1/2$ and $J=3/2$ scattering lengths for
$I=0$ and $I=1$,
\begin{equation}
a^{IJ}_{D(D^*)N}=-\frac{1}{4 \pi} 
\frac{M_N}{\sqrt{s}} \,T^{IJ}_{D(D^*)N \rightarrow D(D^*)N} \ ,
\end{equation}
at $D(D^*)N$ threshold and with $M_N$  the nucleon mass. For the $DN$
effective interaction, we find that our $I=0$ scattering length is
negligible compared to the $I=1$ one, in contrast to the SU(4) TVME
model of Ref.~\cite{lutz} or the meson-exchange model of the J\"ulich
group \cite{haide2}. Moreover, our positive scattering lengths
indicate the attractive behavior of the $D$ meson self-energy close
to threshold in contrast with the values of those previous
references. The discrepancy with previous works has its origin in the
different resonant-hole composition of the self-energy close to
threshold. Moreover, as a new development, we also provide  
the $D^*N$ scattering lengths. We note that
the dominant repulsive contribution comes from the $J=3/2$ partial
wave. 

Calculated scattering lengths come out radically different from those deduced within the Born approximation. This is not surprising because of the strong character of the meson-baryon interaction in the $C=1$ sector, and the existence of resonances, which hint non-perturbative physics, close to the $D^{(*)}N$ threshold (in some cases they are placed even below it).

\begin{figure}[t]
\begin{center}
\includegraphics[width=0.5\textwidth]{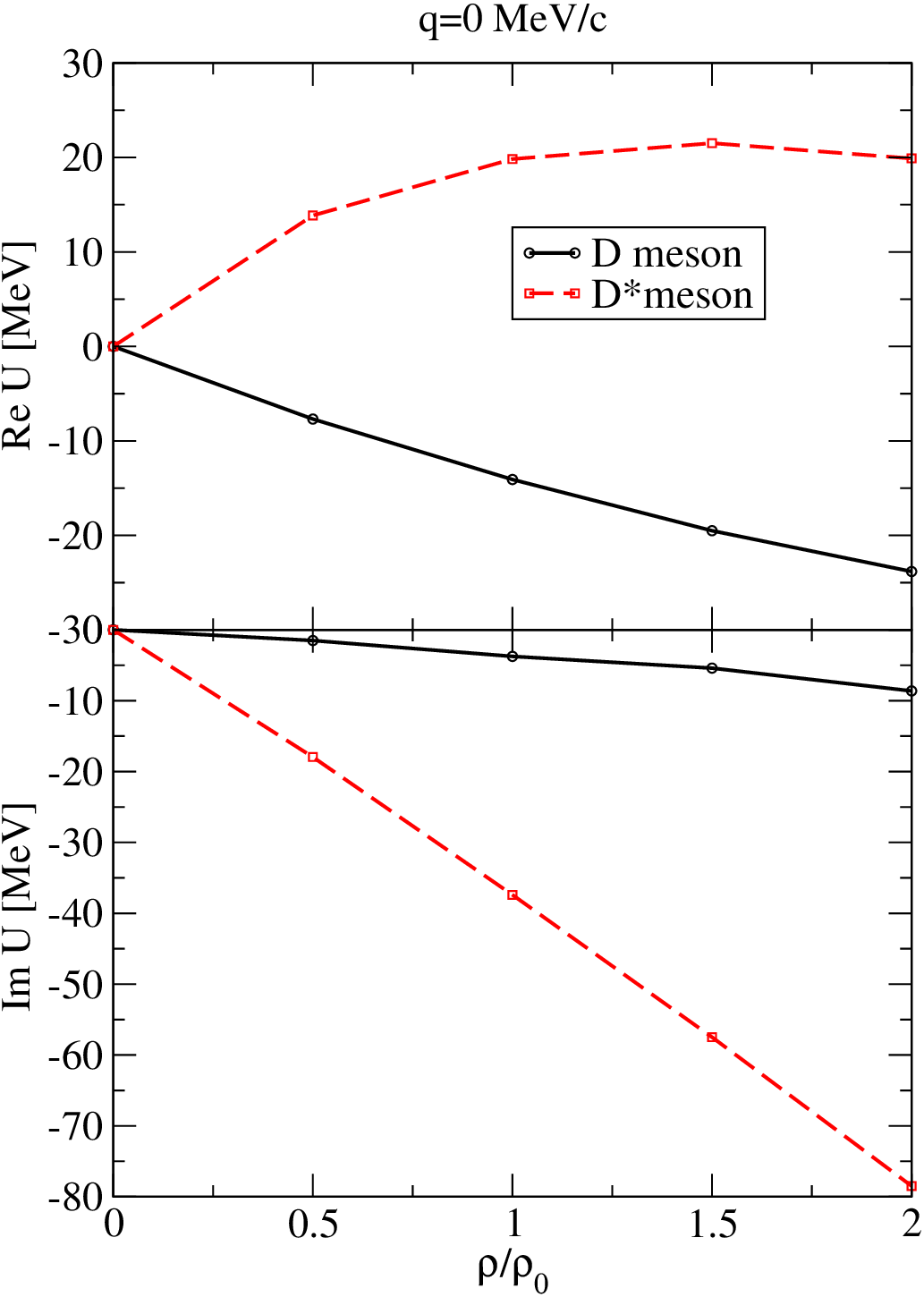}
\caption{$D$ and $D^*$ optical potentials as a function
of $\rho/\rho_0$ for $|\vec{q}|=0 \ {\rm MeV/c}$.  }
\label{fig:upot}
\end{center}
\end{figure}

We can now define the $D$ and $D^*$ optical potentials in the nuclear
medium as
\begin{equation}
U(\vec{q\,})=
\frac{\Pi(\omega_{qp}(\vec{q}\,),\vec{q}\,)}{2 \,\sqrt{m^2+\vec{q}\,^2}} \ ,
\end{equation}
which, at zero momentum, can be identified as the in-medium shift of
the $D$ and $D^*$ meson mass. In
Fig.~\ref{fig:upot} the real and imaginary parts of the optical potential are shown as function of the density for
$|\vec{q}|=0 \ {\rm MeV/c}$. The symbols indicate the calculated values for the optical potentials.
We refrain from showing finite momentum results for the optical potential due to the uncertainties present in our calculation once we move away from threshold as result of the WT interaction together with ignoring higher-multipolarity interactions. 

The mass shift at $\rho=\rho_0$ stays attractive for the $D$ meson
while becomes repulsive for the $D^*$ meson with increasing density,
in correspondence with the behavior of the quasiparticle peaks in
Fig.~\ref{fig:spec}. Similar values for the $D$ meson potential were
obtained for in-medium models
of Refs. \cite{lutz,angels-mizutani,tolos-angels-mizutani,tolos-schaffne-mishra,tolos-schaffne-stoecker}. However,
as explained in Fig.~\ref{fig:spec}, the origin of the attraction can
be traced back to a different resonant-hole contribution in the SU(4)
models compared to the SU(8) case.

The dilution with density of the
spectral functions give rise to the increase of the imaginary part of
the optical potential for both $D$ and $D^*$ mesons, as this corresponds in module to half of the width of the spectral function at the quasiparticle peak. However, the $D$ meson width turns out to be much more smaller in the SU(8) scheme than in the SU(4) models.
Then, we expect to find bound states for the $D^0$-nucleus
system \cite{tsushima}. Looking at the strength of the optical
potential
for density $\rho_0$, we expect those states to be bound at
most by 15~MeV and have half-widths lower than 4~MeV. Hence, we
expect various states to be observable in different  nuclei. It
can be of interest to study the  $D^0$-nucleus spectrum,
binding energies and widths predicted  by the optical potential obtained here
and compare it with predictions of other models.  It is not clear that
the $D^+$-nucleus hadronic attraction will be able to overcome the
Coulomb repulsion to provide similar bound states, that will be a
subject of future research. Experiments to determine such a
$D^0$-nucleus bound states will be welcome.

\section{Conclusions}

We have studied the properties of $D$ and $D^*$ mesons in symmetric
nuclear matter within a simultaneous self-consistent coupled-channel
unitary approach that implements the features of heavy-quark symmetry.
The corresponding in-medium solution incorporates Pauli blocking
effects, and the $D$ and $D^*$ meson self-energies in a
self-consistent manner. In particular, we have analyzed the behavior
of dynamically-generated baryonic resonances in the nuclear medium in
the $C=1$ and $S=0$ sector within this SU(8) spin-flavor symmetric
model and their influence in the self-energy and, hence,the spectral
function of the $D$ and $D^*$ mesons.  We have also obtained the $D$
and $D^*$ scattering lengths, and computed optical potentials for different
 density regimes. We have finally compared our results
with previous SU(4) models
\cite{angels-mizutani,lutz,tolos-angels-mizutani,haide2}, paying a
special attention to the renormalization of the intermediate
propagators in the medium beyond the usual cutoff scheme.

The SU(8) model generates a wider spectrum of resonances with $C=1$
and $S=0$ content compared to the previous SU(4) models. While the
parameters of both SU(4) and SU(8) models are fixed by the ($I=0$,$J=1/2$)
$\Lambda_c(2595)$ resonance, the incorporation of vectors mesons in
the SU(8) scheme generates naturally $J=3/2$ resonances, such as
$\Lambda_c(2660)$, $\Lambda_c(2941)$, $\Sigma_c(2554)$ and
$\Sigma_c(2902)$, which might be identified experimentally
\cite{Yao}. New resonances are also produced for $J=1/2$, as
$\Sigma_c(2823)$ and $\Sigma_c(2868)$, while others are not observed
due to the different symmetry breaking pattern used in both
models. The modifications of the mass and width of these resonances in
the nuclear medium will strongly depend on the coupling to channels
with $D$, $D^*$ and nucleon content. Moreover, the resonances close to
the $DN$ or $D^*N$ thresholds change their properties more evidently
as compared to those far offshell. The improvement in the
regularization/renormalization procedure of the intermediate
propagators in the nuclear medium beyond the usual cutoff method has
also an important effect on the in-medium changes of the
dynamically-generated resonances, in particular, for those lying far
offshell from their dominant channel, as the case of the
$\Lambda_c(2595)$.

The self-energy and, hence, the spectral function of the $D$ and $D^*$
mesons show then a rich spectrum of resonant-hole states. The $D$
meson quasiparticle peak mixes strongly with $\Sigma_c(2823)N^{-1}$
and $\Sigma_c(2868)N^{-1}$ states while the $\Lambda_c(2595)N^{-1}$ is
clearly visible in the low-energy tail. The $D^*$ spectral function
incorporates the $J=3/2$ resonances, and the $\Sigma_c(2902)N^{-1}$
and $\Lambda_c(2941)N^{-1}$ fully mix with the quasiparticle peak. As
density increases, these $\tilde{Y}_cN^{-1}$ modes tend to smear out and the
spectral functions broaden as the collisional and absorption processes
increase. This broadening in dense matter might have important
consequences for the dynamically generation of scalar resonances with
hidden and open charm content \cite{raquel} as well as for excited
charmonium states for the experimental conditions expected in the
PANDA and CBM experiments at FAIR \cite{fair}. This latter
experimental scenario, however, requires the incorporation of finite
temperature effects.

The behavior with density  of the quasiparticle peaks is
better visualized with the optical potentials. The $D$ meson potential
stays attractive while the $D^*$ meson one is repulsive with
increasing density up to twice the normal nuclear
matter one. The attractive and repulsive character of the $DN$ and $D^*N$
interactions close to threshold, respectively, was already observed in
free space via the scattering lengths. In particular, the optical
potentials with density do not follow the low-density approximation,
as expected from the complicated resonant-hole structure of the
self-energy. The imaginary part in both cases increases with
density. Compared to in-medium SU(4) TVME
models, we obtain similar values for the $D$ meson real part of the
potential but much smaller imaginary parts. This result can have
important implications for the observation of $D^0$-nucleus bound
states. Work along this line is in progress.

Future work also includes the study of the influence of the $\Delta$
self-energy in the in-medium $D$ and $D^*$ self-energies as well as
the inclusion of the width of the vector mesons in the meson-baryon
channels. Moreover, finite temperature effects are mandatory for the
analysis and interpretation of the data in the future CBM heavy-ion
experiment at FAIR.

To end up this work, we would like to stress that it would be
important to count with an improved model in the vacuum, and future
research along these lines would be highly desirable. When considering
the findings of this work, one should bear in mind that the
deficiencies of the model in the free space would definitely affect to
results presented here for amplitudes embedded in cold nuclear matter.

\section{Acknowledgments}
C.G.R thanks L.L. Salcedo for useful discussions.  L.T. wishes to acknowledge support from the ``RFF-Open and hidden
 charm at PANDA'' project from the Rosalind Franklin Programme of the
 University of Groningen (The Netherlands) and the Helmholtz
 International Center for FAIR within the framework of the LOEWE
 program by the State of Hesse (Germany). This research is supported by DGI
 and FEDER funds, under contracts FIS2008-01143/FIS and the Spanish
 Consolider-Ingenio 2010 Programme CPAN (CSD2007-00042), by Junta de
 Andaluc\'\i a under contract FQM225. It is part of the European
 Community-Research Infrastructure Integrating Activity ``Study of
 Strongly Interacting Matter'' (acronym HadronPhysics2, Grant
 Agreement n. 227431) and of the EU Human Resources and Mobility
 Activity ``FLAVIAnet'' (contract number MRTN--CT--2006--035482),
 under the Seventh Framework Programme of EU.

%\begin{references}

\end{document}